# Modeling of the dynamic pole-to-pole oscillations of the *min* proteins in bacterial cell division: The effect of an external field

**Charin Modchang, Paisan Kanthang, Wannapong Triampo,* Waipot Ngamsaad, Narin Nuttawut and I-Ming Tang**

Department of Physics and Capability Building Unit in Nanoscience and Nanotechnology, Faculty of Science, Mahidol University, Bangkok 10400, Thailand

**Yongwimol Lenbury**

Department of Mathematics, Faculty of Science, Mahidol University, Bangkok 10400, Thailand

## Abstract

One of the most important steps in the developmental process of the bacteria cell at the cellular level is the determination of the middle of the cell and the proper placement of the septum, these being essential to the division of the cell. In *E. coli,* this step depends on the proteins MinC, MinD, and MinE. Exposure to a constant electric field may cause the bacteria cell division mechanism to change, resulting in an abnormal cytokinesis. To see the effects of an external field e.g., an electric or magnetic field on this process, we have solved a set of deterministic reaction diffusion equations, which incorporate the influence of an electric field. We have found some changes in the dynamics of the oscillations of the *min* proteins from pole to pole. The numerical results show some interesting effects, which are qualitatively in good agreement with some experimental results.

---

*Corresponding author, E-mail: **wtriampo@yahoo.com**; Fax: 662-201-5843

# I. INTRODUCTION

Cell division is the process by which a cell separates into two new cells after its DNA has been duplicated and distributed into the two regions that will later become the future daughter cells. For a successful cell division to take place, the cell has to determine the optimal location of the cell separation and the time to start the cell cleavage. This involves the identification of the midpoint of the cell where the septum or cleavage furrow will form. For *Escherichia coli* and other rod-like bacteria, evidences have accumulated over the past few years which indicate that the separation into two daughter cells is achieved by forming a septum perpendicular to parent cell's long axis. To induce the separation, the FtsZ ring (Z ring), a tubulin-like GTPase is believed to initiate and guide the septa growth by a process called contraction [1]. The Z ring is usually positioned close to the center, but it can also form in the vicinity of the cell poles. Two processes are known to regulate the placement of the division site: nucleoid occlusion [2] and the action of the *min* proteins [3]. Both processes interfere with the formation of the Z ring that determines the division site. Nucleoid occlusion is based on cytological evidence that indicates that the Z ring assembles preferentially on those portions of the membrane that do not directly surround the dense nucleoid mass [4].

The *min* proteins that control the placement of the division site are the MinC, the MinD, and the MinE proteins [3]. Experiments, involving the use of modified proteins show that MinC is able to inhibit the formation of the FtsZ-ring [5]. MinD is an ATPase that is connected peripherally to the cytoplasmic membrane. It can bind to the MinC and activate the function of the MinC [6,7]. Recent studies show that the MinD can also recruit the MinC to the membrane. This suggests that the MinD stimulates the MinC by concentrating the MinC near to its presumed site of activation [8, 9]. MinE provides topological specificity to the division inhibitor [10]. Its expression results in a site-specific suppression of the MinC/MinD action so that the FtsZ assembly is allowed at the middle of the cell but is blocked at the other sites [3].

In the absence of the MinE, the MinC/MinD is distributed homogeneously over the entire membrane. This results in a complete blockage of the Z-ring formation. The long filamentous cells, which are subsequently formed would not be able divide [8, 9, 11, 12]. Using fluorescent labeling, the MinE was shown to attach to the cell wall only in the presence of the MinD [13, 14]. As MinD dictates the location of MinC, the latter would oscillate by itself. This would result in the concentration of the division inhibitor at the membrane on either cell end, alternating between being high or very low every other 20 s or so [8, 9]. The presence of MinE is not only required for the MinC/MinD oscillation, it is also involved in setting the frequency of the oscillation cycle [11]. Several sets of evidence indicate that the MinE localization cycle is tightly coupled to the oscillation cycle of MinD.

Recent microscopy of the fluorescent labeled proteins involved in the regulation of *E. coli* division have uncovered stable and coherent oscillations (both spatial and temporal) of these three proteins [15]. The proteins oscillate from one end to the other end of the bacterium, moving between the cytoplasmic membrane and cytoplasm. The detail mechanism by which these proteins determine the correct position of the division plane is currently unknown, but the observed pole-to-pole oscillations of the corresponding distribution are thought to be of functional importance. Under different culture conditions and/or environment changes, (e.g., pH, light, and external field) changes in the pole-to-pole oscillations could affect the growth of the bacteria. Here we discuss only the effects of an electric field.

In the present work, we use a mathematical approach to investigate the influence of the external constant external field on the cytokinesis mediated by *min* protein pole-to-pole oscillation. We propose a mathematical model and then solve it numerically to see how the *min* protein oscillation mechanism for the bacteria cell division may change. We also present some comments about the connection between our mathematical approach and the real world experimental results.

## II. Model

Sets of reaction-diffusion equations have often been used in biological applications to model self-organization and pattern formation [16]. These mathematical equations have two components. The first component is the diffusion term that describes the diffusion of the chemical species. At the molecular level, the diffusion term often results in a net flow of chemical species from the region of high concentration to regions of lower concentration. The second component is the reaction term that describes the self-organization of the biological systems.

We have adopted the dynamic model of compartmentization in the bacterial cell division process proposed by Howard *et. al*. [17] by adding an extra term that depend on the external electric fields. The dynamics of the bacteria in the presence of the external filed is described by a set of four non-linear coupled reaction-diffusion equations. We focus on the *E. coli* bacteria, which is a commonly studied rod shaped bacteria of approximately $2-6\mu m$ in length and around $1-1.5\mu m$ in diameter. *E. coli* divides roughly every hour via cytokinesis. Our starting point is the set of one dimensional deterministic coupled reaction-diffusion equations describing the dynamics of the interactions between the local concentration of the MinD and MinE proteins. The equations describe the time rates of change of the concentration due to the diffusion of the MinD and the MinE and to the transfer between the cell membrane and the cytoplasm. The dynamics of these *min* proteins in the presence of an external field, are described by:

$$\frac{\partial \rho_D}{\partial t} = D_D \frac{\partial^2 \rho_D}{\partial x^2} + J_D \frac{\partial \rho_D}{\partial x} - \frac{\sigma_1 \rho_D}{1+\sigma_1' \rho_e} + \sigma_2 \rho_e \rho_d \ , \qquad (1)$$

$$\frac{\partial \rho_d}{\partial t} = D_d \frac{\partial^2 \rho_d}{\partial x^2} + J_d \frac{\partial \rho_d}{\partial x} + \frac{\sigma_1 \rho_D}{1+\sigma_1' \rho_e} - \sigma_2 \rho_e \rho_d \, , \qquad (2)$$

$$\frac{\partial \rho_E}{\partial t} = D_E \frac{\partial^2 \rho_E}{\partial x^2} + J_E \frac{\partial \rho_E}{\partial x} - \sigma_3 \rho_D \rho_E + \frac{\sigma_4 \rho_e}{1+\sigma_4' \rho_D} \, , \qquad (3)$$

and

$$\frac{\partial \rho_e}{\partial t} = D_e \frac{\partial^2 \rho_e}{\partial x^2} + J_e \frac{\partial \rho_e}{\partial x} + \sigma_3 \rho_D \rho_E - \frac{\sigma_4 \rho_e}{1 + \sigma_4' \rho_D} \tag{4}$$

where $\rho_D, \rho_E$ are the concentrations of protein MinD and MinE in the cytoplasm, respectively. $\rho_d, \rho_e$ are the concentrations of protein MinD and MinE on the cytoplasmic membrane. The first equation describes the time rate of change of the concentration of MinD $(\rho_D)$ in the cytoplasm. The second is for the change in the MinD concentrations $(\rho_d)$ on the cytoplasmic membrane. The third is for the change of the concentration of MinE $(\rho_E)$ in the cytoplasm. The last one is for the change in the MinE concentrations $(\rho_e)$ on the cytoplasmic membrane. Since the experimental results given in [9], show that the MinC dynamics simply follows that of the MinD, we have not written out the equations for the MinC explicitly.

The importance feature of our model is the second terms on the right hand sides of the equations. They represent the effect of the external field in the reaction-diffusion equation [18, 19] controlled by the external field parameter. We assume that the chemical substance moving in the regions of an external field will experience the force that is proportional to the external field parameter $J$ times the gradient of the concentration of that substance. In general $J = \mu E$ where $E$ is the field strength and $\mu$ is the ionic mobility of the chemical substance. $\mu$, in general, will be proportional to the diffusion coefficient of the chemical substance and will depend on the total amount of free charge in that substance. In this model $J_i = \mu_i E \quad \{i = D, E, d, e\}$ is the external field parameter for each protein types.

We assume that the diffusion coefficients $(D_D, D_d, D_E, D_e)$ are isotropic and independent of $x$. The constant $\sigma_1$ represents the association of MinD to the membrane wall [12]. $\sigma_1'$ corresponds to the membrane-bound MinE suppressing the recruitment of MinD from the cytoplasm. $\sigma_2$ reflects the rate that the MinE on the membrane drives the MinD on the membrane into the cytoplasm. Based on the evidence of the cytoplasmic interaction between MinD and MinE [7], we let $\sigma_3$ be the rate that cytoplasmic MinD recruits the cytoplasmic MinE to the membrane while $\sigma_4$ corresponds to the rate of dissociation of MinE from the membrane to the cytoplasm. Finally, $\sigma_4'$ corresponds to the cytoplasmic MinD suppressing the release of the membrane-bound MinE. Evidence points to most of the diffusion process occurring in the cytoplasm. It is therefore reasonable to set $D_d$ and $D_e$ to zero. It follows immediately that $\mu_d = \mu_e = 0$ and so $J_d = J_e = 0$

In our model we assume that the total number of each type of protein is conserved. We further assume that the *min* proteins can bind/unbind from the membrane and that the proteins do not degrade during the process. The zero flux boundary conditions are imposed at both ends of the bacterium. The total amounts of MinD and MinE, obtained by integrating $\rho_D + \rho_d$ and $\rho_E + \rho_e$ over the length of the bacterium, are conserved.

## III. Numerical results and discussion

Since the bacterium length is very short, it is reasonable to assume that the applied electric field has a constant value throughout the bacterium length. We have numerically solved the set of four coupled reaction-diffusion equations (1)-(4) by using the explicit Euler method [20]. The size of *E. coli* is taken to be $2\mu m$ in length. The total time needed for each simulation is approximately $10^4$ s. In our simulations we have discretized space and time, i.e., we have taken $dx = 8\times10^{-3}\mu m$ and $dt = 1\times10^{-5}s$. The space covering the bacteria will be divided into 251 grid points and the time has been divided into $10^9$ times steps ($10^9$ iteration steps). Initially we assume that the MinD and MinE are mainly at the opposite ends of the bacterium with the number of *min* molecules in each cell being 3000 for the MinD population [6] and 170 for the MinE population [21]. Since the total amount of MinD and MinE in *E.coli* must conserve, we have set the flux of MinD and MinE to be zero at both ends of the bacterium. Since there is no experimental values of $\mu$ for either MinD and MinE, we work with the external field parameter $J$ , which is proportional to *E*, instead of *E* explicitly. We also assume that $\mu_D = \mu_E$ (we assume MinD and MinE have the same type of charges). It follows immediately that $J_D = J_E \equiv J$ . The values of the other parameters are: $D_D = 0.28\mu m^2 s^{-1}$, $D_E = 0.6\mu m^2 s^{-1}$, $\sigma_1 = 20s^{-1}$, $\sigma'_1 = 0.028\mu m$, $\sigma_2 = 0.0063\mu m s^{-1}$, $\sigma_3 = 0.04\mu m s^{-1}$, $\sigma_4 = 0.8s^{-1}$, and $\sigma'_4 = 0.027\mu m$ . In our analyses of the numerical results, we looked at the time-averaged values of the concentrations of MinD and MinE and at the patterns of the oscillations of MinD and MinE at various *J* values.

In the absent of the external field, the numerical results [17] show that most of the MinD will be concentrated at the membrane and the MinE at the midcell. This would result in an accurate division at the midcell. In the presence of the external field, both MinD and MinE will experience the force in the same direction. This force causes a shift of the time average minimum of MinD. This would shift the division site from being at the midcell. Our numerical solutions show that the behavior of the Min system in the presence of an external field will depend on the strengths of the external field parameter $(J)$.

Figure 1 shows the oscillation patterns for $J_E = J_D \equiv J = 0.0$ m/s to *J* =0.4 m/s. It is seen that as *J* increase, both the MinD and MinE concentrations in the left part of the *E.coli* becomes larger while in the right part, the two concentrations become smaller as *J* is increased. This behavior is a reflection of the fact that the external force is acting in the left direction. These patterns show the shifting in the concentrations of the *min* proteins towards the left pole.

In Figure 2 we show the time-averaged concentrations of the MinD and MinE proteins at different positions within the bacteria. In these curves, positive values of the external field parameter are used. From this Figure, we see that in the case of no external field ($J$= 0.0) the time-averaged concentrations of MinD and MinE are symmetric about the midcell. MinD has a minimum at the midcell while MinE has a maximum at the midcell. When an external field is applied, we see a shift in the minima of MinD and in the maxima of MinE. The time-averaged concentration curves are no longer symmetric about the midcell. In nature, the MinE protein looks like a ring structure that effectively positions the anti-MinCD activity [14, 11]. MinCD inhibits the division process, so in nature the bacterium divides at the site where the minimum MinD concentration occurs. The value of the MinE concentration is not maximum at the midcell. The minimum of the MinD shifts to right pole under the influence of a positive $J$ values.

We have measured the percentage of shifting of the time-averaged concentration in the local minima of the MinD and local maxima of the MinE. This is shown in Figure 3. The figure shows that the minimum of MinD is always shift to the right pole. This is the result of the external force pulling the MinD to the left. The maximum of MinE is not always shift to the right. When $J$ < 0.2 m/s the maximum of MinE is shifted to the right but when $J$ > 0.2 it shifted to the left of the midcell. This difference arises because of the relative magnitudes of the forces acting on the two proteins. First of all, there is an internal force between the MinD and MinE proteins. This force causes the MinE to repel the MinD. In the absence of any other forces, this would explain why the location of the maximum of MinE would be the location of minimum of MinD. When an external field is applied (as expressed by a non zero value of J), then one must take into account the relative magnitudes of the two forces.

When $J$ is large (larger than 2 m/s) the external force would dominate the internal force between the MinD and MinE proteins. The external force would pull the MinD and MinE in the same direction causing the location of the maximum of the MinE to be no longer at the location of the minimum of the MinD. If $J$ is small (smaller than 0.2 m/s), the internal force between MinD and MinE will be dominate.

This would result in the two location (one of the maximum of MinE and the other of the minimum of MinD) to be nearly the same. In Fig. 3, we also see that the shifts of the minimum of the MinD concentrations increase as the field parameter $J$ increase. Since the division site will be the location where MinD concentration is minimum, the shifting in the minimum of MinD concentration to the right pole indicate that the division site must also sift to the right pole. When we set $J$ to be the negative, the results are very similar to those of the positive $J$ values as expected, curve for the time averages of the concentration of the *min* proteins shifts in the mirror side about the midcell.

In Figs. 4a and 4b. we show the concentrations of the MinD and MinE proteins at the left end grid, the middle grid and the right end grid versus time. In these figures, it is easy to see that when $J$= 0.0 m/s, the concentration of MinD (or MinE) at the left end grid and the right end grid have the same pattern of oscillation, with the same frequency and amplitude, with phase difference 180º. At the midcell grid, the frequency of the oscillation is two times greater than that of right end grid. When the external field is applied, the amplitude of the oscillations at the two end grids are no longer equal but the frequency of the oscillation of the three grids become the same. As $J$ is increased, the amplitude of the oscillation at the right end grid is seen to decrease while that of the left end and midcell grids are seen to increase.

Figure 5 show the periods of the oscillation of MinD concentration at the left end grid for various value of $J$. In this figure we see that for the case of no external field, the period of the oscillation is equal to 115 s which is in good agreement with the experimental value. When the external field is applied, the period of the oscillation is seen to increase. When $J$ is not too large ($J$<0.3) the period of the oscillation will increase as $J$ is increased. The increase in the period of oscillation as an external field is applied indicates that in the presence of an external field, the bacterium needs a longer time to divide.

## IV. CONCLUDING REMARKS

Proper divisions of the bacteria require the accurate definition of the division site [3]. This accurate identification of the division site is determined by the rapid pole-to-pole oscillations of the MinCDE [8, 11, 22]. Using a mathematical model to describe the dynamics of the *min* pole-to-pole oscillations, Howard *et al.* [17], found that the midcell position in the *Escherichia coli* bacteria, correspond to the point where the time averaged MinD and MinE concentration were minimum and maximum, respectively. They also found that the concentrations of these two proteins were symmetric about the midcell position.

To see the effect of exposing a *E. coli* bacteria to an electric field, we have added some additional terms to the reaction diffusion equations for the pole-to-pole oscillation of the *min* proteins in the *E. coli* bacteria proposed by Howard *et al.* The additional terms are the gradient terms appearing in eqns. (1)–(4). These terms depend on the strength of the external field and the charge of the protein. We then used a numerical scheme to solve the resulting coarse-grained coupled reaction-diffusion equations. The results are shown in Figures 1 to 5. Our results shows deviations from the results obtained by Howard *et al.*, e.g. the concentrations of the MinD and MinE are no longer symmetric about the middle of the long axis nor are the minimum and maximum of the MinD and MinE concentrations at the middle of long axis. The shifting in the minimum of the time average concentration of MinD from the midcell should results in the shifting of the division site. The shift of the minimum concentration of MinD from the mid point appears to be dependent on the strength of the external field. This indicates that if the parent cell can divide under these condition it must divide into two filamentous cells, provide that the external field is strong enough. Since the external field can cause the shifting in the minimum of the time average concentration of MinD, an external electric field can interfere with the division process.

## Acknowledgements

We thank M. Howard, J. Wong-ekkabut and M. Chooduang for their useful comments and suggestion. This research is supported in part by Thailand Research Fund through the grant number TRG4580090 and RTA4580005. The IRPUS Program 2547 to Charin Modchang and W. Triampo is acknowledged. The Development and Promotion of Science and Technology talents program to Waipot Ngamsaad.

# Figures

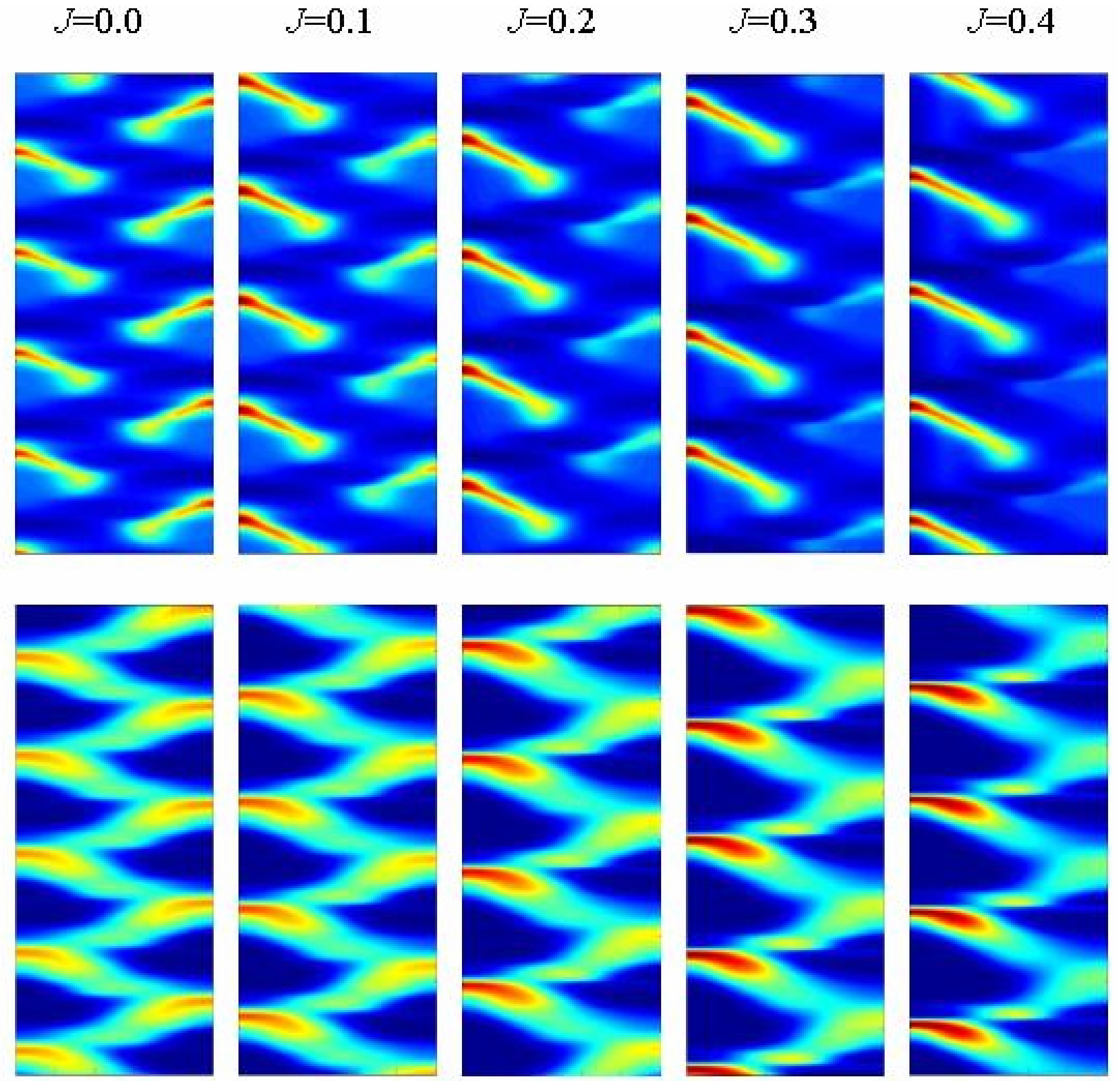

FIG. 1. Space-time plots of the total $(\rho_D + \rho_d)$ MinD (above) and total $(\rho_E + \rho_e)$ MinE (below) concentration for $J = 0.0$ m/s to $J$ = 0.4 m/s. The color scale, runs from blue to red, denotes an increasing in the concentration from the lowest to the highest. The MinD depletion from midcell and the MinE enhancement at the midcell are immediately seen. The vertical scale spans time for 500 s. The times increase from bottom to top and the oscillations pattern repeats infinitely as time increases. The horizontal scale spans the bacterial length ($2\,\mu m.$). Note the increase in the MinD and MinE concentrations at the left end of the bacterium as $J$ increases.

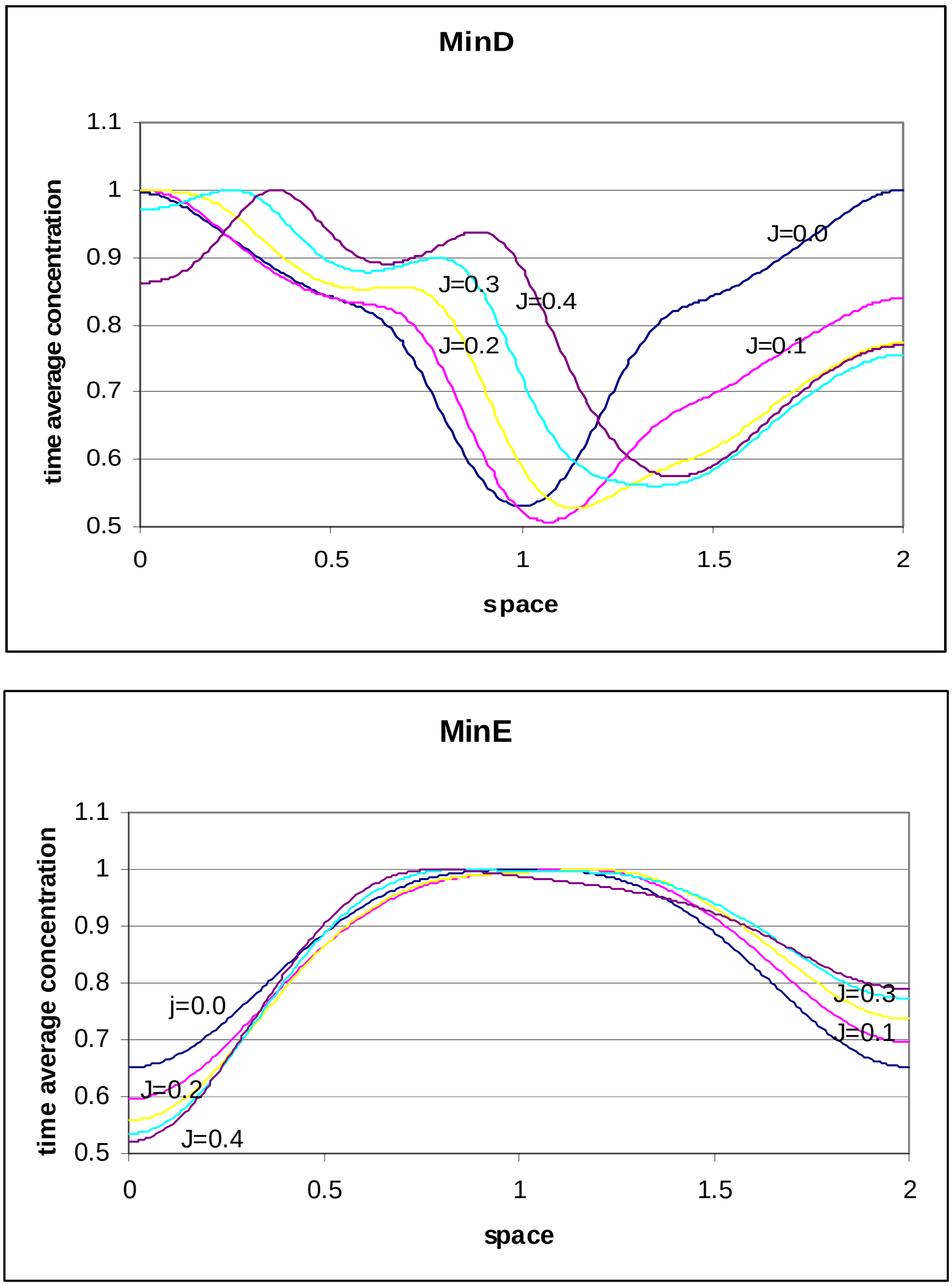

FIG. 2. The time average concentration of MinD (above) and MinE (below) relative to their respective time-averaged maxima, $\langle \rho(x) \rangle / \rho_{\max}$, as a function of position $x$ (in $\mu m$) along the bacterium axis under the influence of positive values of the static external field. The curves show a shift in the local minima of the MinD and the local maxima of the MinE from the midcell depending on the strength of the field.

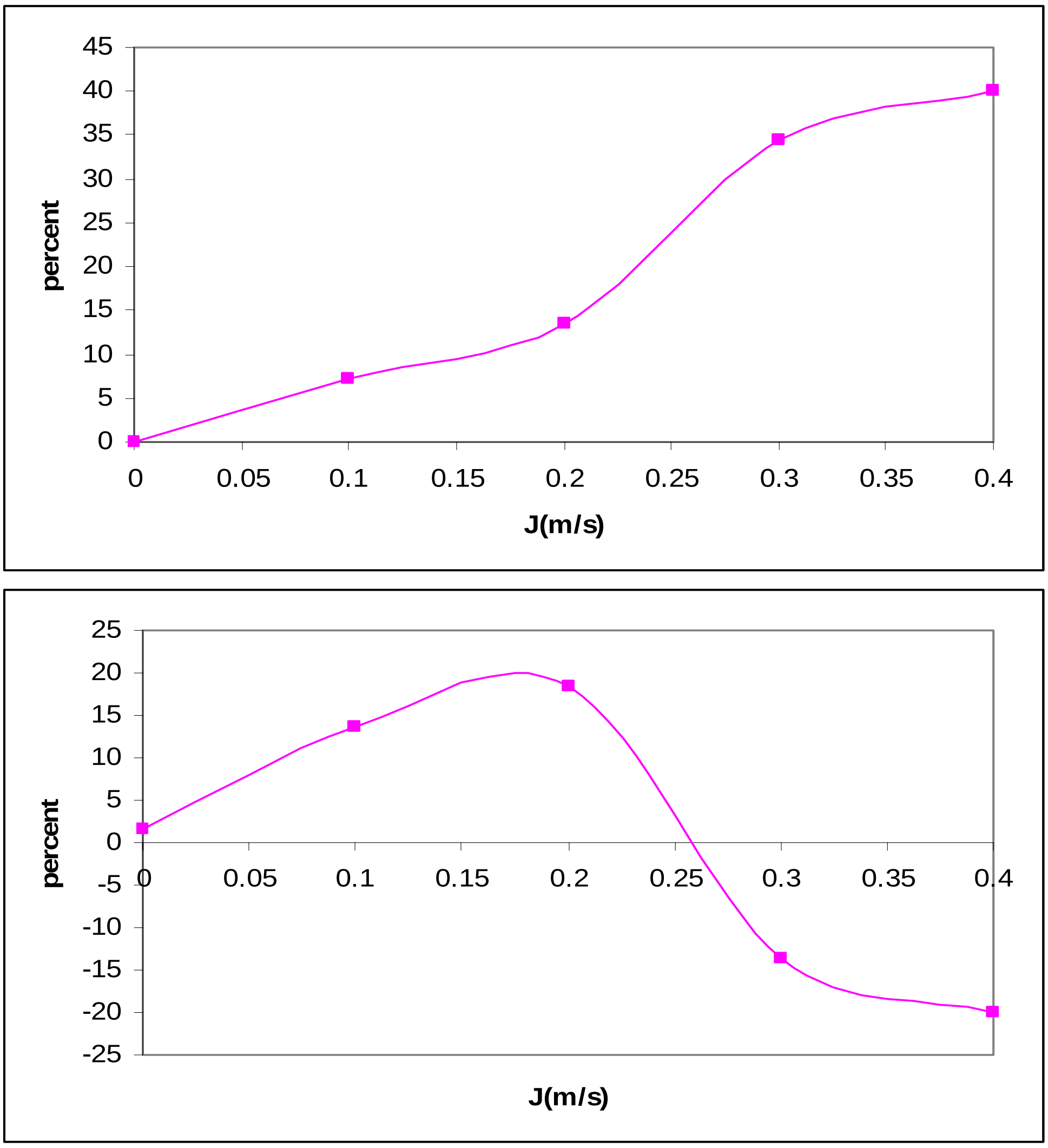

FIG. 3. The percentages of the shifting of the local minima of MinD (above) and the local maxima of MinE (below) from the midcell at the various values of *J*. Positive values denote the shifting to the right pole and negative value to the left pole.

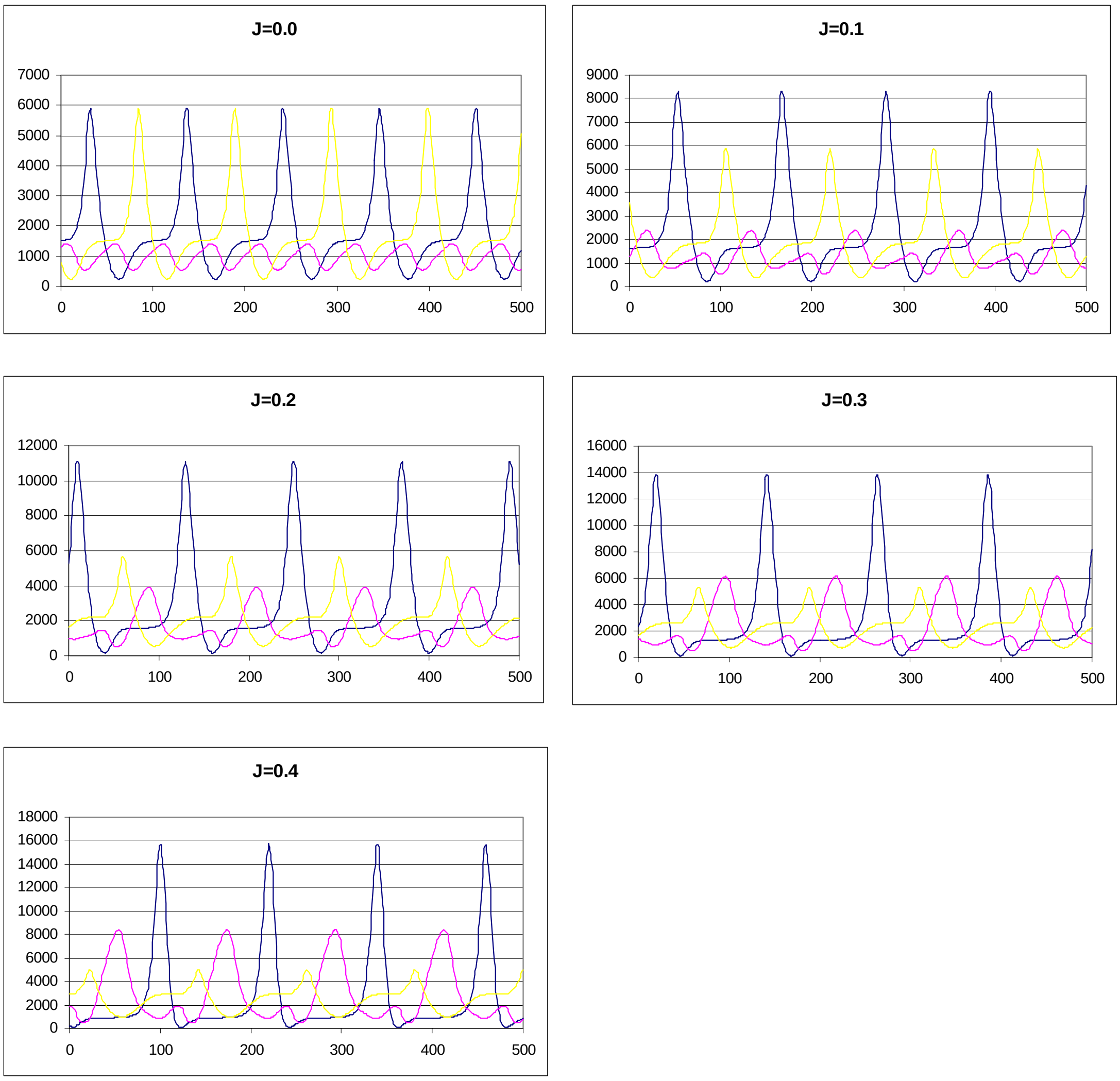

FIG. 4a. Plots of the concentration of MinD at the left end grid (blue), the middle grid (pink) and the right end grid (yellow) versus time in seconds for $J$ = 0.0 m/s to $J$ = 0.4 m/s. The verticals scale span for concentration in molecule per meter.

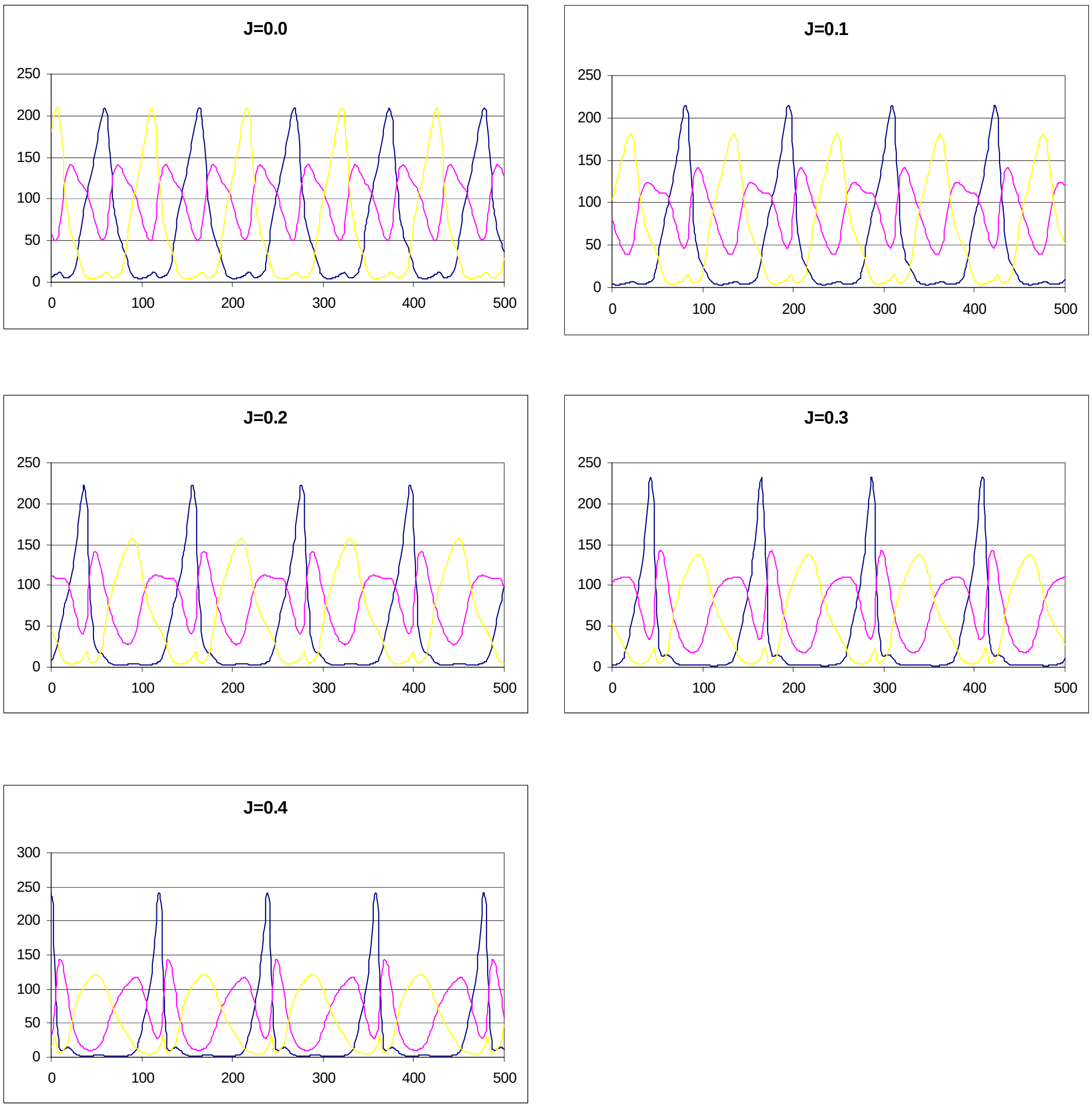

FIG. 4b. Plots of the concentration of MinE at the left end grid (blue), the middle grid (pink) and the right end grid (yellow) as a function of time in seconds for $J$ = 0.0 m/s to $J$ = 0.4 m/s. The verticals scale span for concentration in molecule per meter.

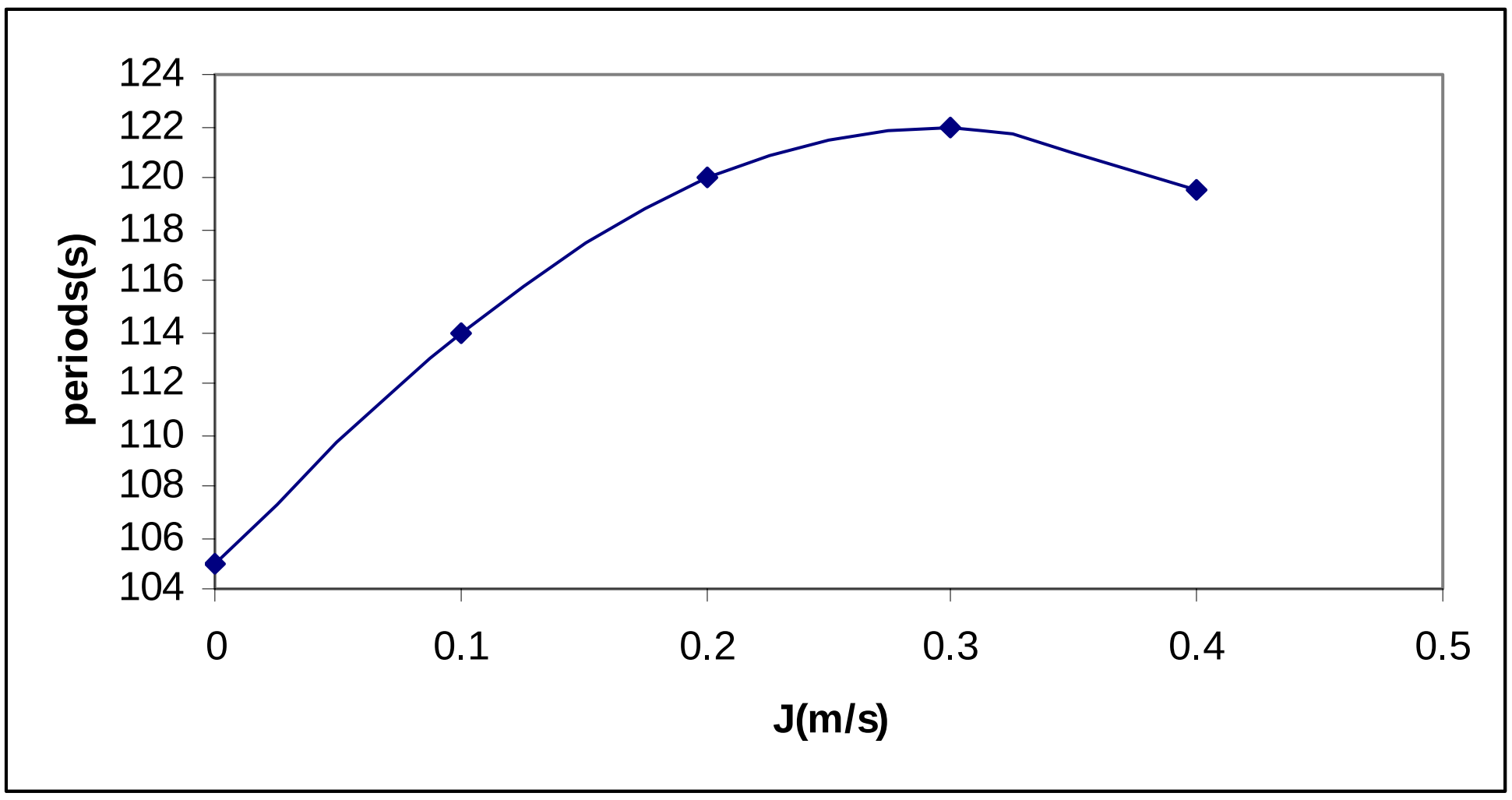

FIG. 5. The periods of the oscillation of the MinD concentration at the left end grid at the various values of *J*. The curve show the increasing in the period of oscillation as *J* increase, indicate that the bacterium would spend more time to divide.